\documentclass[sn-mathphys,Numbered]{sn-jnl}

\usepackage{graphicx}%
\usepackage{multirow}%
\usepackage{amsmath,amssymb,amsfonts}%
\usepackage{amsthm}%
\usepackage{mathrsfs}%
\usepackage[title]{appendix}%
\usepackage{xcolor}%
\usepackage{textcomp}%
\usepackage{manyfoot}%
\usepackage{booktabs}%
\usepackage{algorithm}%
\usepackage{algorithmicx}%
\usepackage{algpseudocode}%
\usepackage{listings}%

\theoremstyle{thmstyleone}%
\theoremstyle{thmstyletwo}%

\theoremstyle{thmstylethree}%

\begin{document}

\title[Article Title]{Exploring the Intersection of Complex Aesthetics and Generative AI for Promoting Cultural Creativity in Rural China after the Post-Pandemic Era}


\author[1]{\fnm{Mengyao} \sur{Guo}}\email{yc07330@um.edu.mo}
\equalcont{These authors contributed equally to this work.}

\author[2,6]{\fnm{Xiaolin} \sur{Zhang}}\email{zhangxiaolin@pku.edu.cn}
\equalcont{These authors contributed equally to this work.}

\author[3]{\fnm{Yuan} \sur{Zhuang}}\email{202020162@mail.sdu.edu.cn}

\author[2,4]{\fnm{Jing} \sur{Chen}}\email{cjing@pku.edu.cn}

\author[5]{\fnm{Pengfei} \sur{Wang}}\email{wangziyunlp@gmail.com}

\author*[7]{\fnm{Ze} \sur{Gao}}\email{zgaoap@connect.ust.hk}

\affil[1]{\orgdiv{}, \orgname{University of Macau}, \orgaddress{\city{Macau}, \postcode{999078}, \country{China}}}

\affil[2]{\orgdiv{}, \orgname{Peking University}, \orgaddress{\city{Beijing}, \postcode{100091}, \country{China}}}

\affil[3]{\orgdiv{}, \orgname{Shandong University}, \orgaddress{\city{Jinan}, \postcode{250100}, \country{China}}}

\affil[4]{\orgdiv{}, \orgname{China Agricultural University}, \orgaddress{\city{Beijing}, \postcode{100091}, \country{China}}}

\affil[5]{\orgdiv{}, \orgname{Tsinghua University}, \orgaddress{\city{Beijing}, \postcode{100091}, \country{China}}}

\affil[6]{\orgdiv{}, \orgname{University of Auckland}, \orgaddress{\city{Auckland}, \postcode{1010}, \country{New Zealand}}}

\affil*[7]{\orgdiv{}, \orgname{Hong Kong University of Science and Technology}, \orgaddress{\city{Hong Kong SAR}, \postcode{999077 }, \country{China}}}
\equalcont{These authors contributed equally to this work.}


\abstract{

This paper explores using generative AI and aesthetics to promote cultural creativity in rural China amidst COVID-19's impact. Through literature reviews, case studies, surveys, and text analysis, it examines art and technology applications in rural contexts and identifies key challenges. The study finds artworks often fail to resonate locally, while reliance on external artists limits sustainability. Hence, nurturing grassroots “artist-villagers” through AI is proposed. Our approach involves training machine learning on subjective aesthetics to generate culturally relevant content. Interactive AI media can also boost tourism while preserving heritage. This pioneering research puts forth original perspectives on the intersection of AI and aesthetics to invigorate rural culture. It advocates holistic integration of technology and emphasizes AI’s potential as a creative enabler versus replacement. Ultimately, it lays the groundwork for further exploration of leveraging AI innovations to empower rural communities. This timely study contributes to growing interest in emerging technologies to address critical issues facing rural China.
}




\maketitle

\section{Introduction}\label{sec1}

With the COVID-19 pandemic devastating rural areas of China, especially tourism, there is an urgent need for innovative approaches to promote economic development and cultural creativity in these regions~\cite{li2022research}. This paper explores the potential of leveraging generative artificial intelligence (AI) in conjunction with complex aesthetics to foster cultural creativity in rural China during the post-pandemic period. Through a literature review, case studies, and data analysis, this research aims to bridge the gap in understanding how machine learning algorithms can be used to generate culturally creative outputs when provided with subjective aesthetic data.

The paper begins by establishing the research background, underscoring how the pandemic has impacted rural Chinese communities reliant on tourism. It then identifies a research gap, noting that while generative AI shows promise in this context, machine learning excels at reproducing technical qualities rather than nuanced aesthetic values\cite{oh2020understanding}. Hence, human input is vital for imparting artistic inspiration to machine learning\cite{hitsuwari2023does}. Next, the paper outlines its contribution - presenting a novel framework utilizing generative AI and complex aesthetics to stimulate cultural creativity.

The study's structure is elaborated, explaining that it employs a mixed methods approach with case studies, a questionnaire, interviews, and text analysis. Contemporary art cases are reviewed, providing examples of art intersecting with rural landscapes. Early examples of using art for rural revitalization in international and Chinese contexts are also summarized. Additionally, existing instances of art and technology collaborating in rural settings are discussed.

Key findings are highlighted, underscoring the importance of site-specific artworks organically woven into local contexts and the potential for generative AI to nurture grassroots artists within villages. In conclusion, this study offers an original perspective on leveraging generative AI and aesthetics to boost cultural creativity in rural China during the post-pandemic recovery. It provides a foundation for further research at the intersection of technology and culture.

\subsection{Research Background}

The primary objective of this paper is to explore the potential implications of the intersection of complex aesthetics and generative AI for promoting cultural creativity in rural areas of China during the post-pandemic era~\cite{banks2021plague}. The COVID-19 pandemic has devastated rural areas in China, especially in the tourism industry~\cite{song2022does}. As a result, there is an urgent need for innovative approaches to promote economic development and cultural creativity~\cite{gao2022ai}. This paper argues that leveraging the intersection of complex aesthetics and generative AI can provide a promising solution to address this need and promote cultural creativity in rural China~\cite{guo2022coupling}.

\subsection{Research Gap}

The exploration of using generative AI to foster cultural creativity in rural China presents an intriguing opportunity but faces key challenges. While generative AI shows promise for creating novel cultural content, its machine learning algorithms excel more at reproducing technical qualities than nuanced, subjective aesthetics. Additional human inputs are thus critical for imbuing AI systems with artistic inspiration reflective of local cultural contexts~\cite{xie2022marketing}. This research aims to bridge this gap by developing methodologies to incorporate subjective aesthetic data into generative AI models, enabling outputs that authentically enhance rural cultural experiences. However, a review revealed a lack of documented applications of generative AI tailored specifically for rural creativity. Despite the potential benefits for rural communities, this absence represents a missed opportunity. Given generative AI’s aptitude for human-guided synthesis, exploring its integration with rural culture merits investigation. This pioneering research can fill the void by proposing techniques to harness generative AI’s capabilities for rural-focused cultural innovations that preserve heritage and empower communities~\cite{che2018tourism}.

\subsection{Contribution}

This paper puts forth an original framework utilizing generative AI and complex aesthetics to promote cultural creativity in rural China. Our approach is two-pronged: first, harnessing generative AI's capacity for machine learning and data synthesis to produce novel cultural content reflective of local aesthetics. Second, incorporating nuanced human aesthetic perspectives to imbue the machine-generated content with richness, specificity, and authenticity. This framework represents a pioneering attempt to synergize technological and humanistic approaches to inject new vitality into rural cultural heritage. Beyond proposing a workflow, this research aims to illuminate the current realities of rural communities in China and envision future possibilities for their revitalization. By elucidating the intersection of technology and culture, we seek to foster enlightening dialogues about preserving local traditions while embracing modernity. Our ultimate contribution is presenting a human-centered model of technological integration that elevates rather than diminishes indigenous cultural creativity.

\section{Structure of the Research}


This study employs a mixed methods approach incorporating qualitative and quantitative techniques to investigate using generative AI and aesthetics for rural cultural creativity. Contemporary art cases situated in rural areas are first reviewed for contextual examples. Early precedents of art applications for rural development internationally and in China are then summarized. Existing collaborations between art and technology in rural settings are also discussed. The core methodology involves three components: case studies of specific Chinese rural art festivals, a questionnaire surveying various stakeholders, and semi-structured interviews with key curators for expert insights. Additionally, textual data is thematically analyzed to identify trends. Together, this multi-faceted approach enables examining the intersection of generative AI, aesthetics, and rural creativity from theoretical and practical lenses. It provides a detailed investigation from reviewing relevant examples to gathering first-hand perspectives. Ultimately, the analysis extracts key findings to inform a proposed framework for applying this intersection to enhance creative outputs. This comprehensive methodology combines literature, cases, surveys, interviews, and textual analysis to holistically explore the potential of generative AI and aesthetics in promoting rural cultural creativity.


\section{Related Works}

Through artwork events, exhibitions, and exchanges, rural areas can revitalize and gain regional developments by having a higher potential to interact with the crowds and face the market on a larger scale ~\cite{saxena2008integrated}. In this digital new era, content creation (such as NFTs) with AI~\cite{wang2022decentralized} gives people a great deal of creative space and imagination, especially artists and writers~\cite{rezwana2023towards}~\cite{sun2023inspire}. The ways people enjoy rural cultural heritage sites and countryside spots have significantly evolved with the technology developments. Visitors can not only be allowed to look and read but also touch, interact, and even create~\cite{pisoni2021human}~\cite{gao2021intelligent}. The following related works chronologically trace the development of the application of complex aesthetics and generative AI in cultural experiences and local heritage. It includes the contemporary art cases that happen on non-urban landscapes, the earlier examples internationally and domestically of the art applications for rural revitalization, and those art and technology intervening in rural areas.

\subsection{The most earlier examples internationally of the art applications for rural revitalization}

 Agnes Denes's ``The Tree Mountain" (1996) was completed by tens of thousands of people from all over the world planting eleven thousand trees in the Pincio gravel pit near Ilojavi, Finland, and eventually formed a huge artificial mountain that became part of the artist's land reclamation project. The project's significance is to relieve ecological pressure on the area by planting trees, with the expectation that a primary forest will eventually be created~\cite{wildysculpted}. Another example South Korea's ``Gamcheon Culture Village" (2009) was built in the 1920s and 1930s, and in 2009 the local authorities, under the Empty Homes Residential Preservation Project, transformed the village into a cultural center with art installations as the primary form of presentation and colorful houses. The value of the project is that it improves the poverty and poor living conditions in the area and provides an example of artistic village revitalization~\cite{choi2018reinvented}. 

These international examples demonstrate the broader significance of art as a catalyst for rural rejuvenation and community engagement. The integration of artistic interventions in rural areas preserves cultural traditions and addresses environmental, social, and economic challenges, making a remarkable impact on both local and international levels~\cite{crespi2007meaning}.

\subsection{The most earlier examples domestically of the art applications for rural revitalization}

The ``Tongling Countryside Public Art Festival'', organized by Kegon Leung in 2018, brings together artists, architects, poets, etc., combining theatre, design, installation, and other art forms. The project has not only made Plough Bridge Village the focus of rural tourism but has also driven the development of folklore, culture, and leisure industries, making it more profitable for villagers~\footnote{Tongling Countryside Public Art Festival: https://www.sohu.com/a/514768925\_121106991}. The ``Daojiao New Art Festival'', initiated by Li Zhenhua and others, is a fusion of new media art and technology that transforms industrial parks and grain silos into festival venues. The festival has helped local people to gain a better understanding of contemporary art, creating a new industrial system and cityscape, as well as boosting the local economy through the creation of an art town~\footnote{Daojiao New Art Festival: https://www.sohu.com/a/145276407\_340420}. The Lianzhou International Foto Festival annually presented a unique theme since its inception in 2005, showcasing diverse Chinese images and exploring global photography. With over 80 exhibitions and its participation from top international photographers and scholars, it is one of China's most influential international photography festivals~\cite{soutter2018notes}.

These rural revitalization projects in China have played an essential role in national rural revitalization efforts. However, so far, only one of these projects has maintained its artistic form to this day, and the Daojiao New Art Festival has only been held once. Meanwhile, the once glorious Tongling Countryside Public Art Festival has gradually lost its former splendor. These few cases, on the one hand, have provided a model for revitalizing rural areas through art in China. Still, on the other hand, they have also exposed the shortcomings in its implementation process.

\subsection{Latest application using AIGC tools to boost rural tourism and for rural revitalization }

The ``Yunshang Ethnic Villages" project launched in May 2023 uses advanced generative AI to recreate rural cultural heritage, weaving community narratives into interactive 3D simulations. Digital technology provides new ways of information dissemination and brand-new experience of traveling experiences. Visitors can quickly obtain detailed information on rural tourist attractions, folk culture, and specialty products through virtual reality technology, intelligent portal websites, live broadcast tools, etc. AIGC-powered products are also shown in this festival, aiming to gain more tourists' attention and for inheritance purposes.

30,000 sample images were collected to integrate and extract the cultural characteristics of the She nationality with the help of Boundless AI. Key cultural elements were digitalized and modeled to generate unique She nationality architecture, character images, clothing, and cultural products innovatively. The AIGC-endowed capability of cultural symbol extraction and content generation significantly boosts ethnic minority village cultural derivatives and digital collections, such as symbols and artworks.


With the development of artificial intelligence, new energy, 5G, and other technologies, rural revitalization breaks through the traditional paradigm with the help of technological means to lead rural change~\cite{sarangi2019survey}. The ``Safe Countryside AI Empowerment" project in Baoshan, Yunnan Province, provides AI technology to address social governance issues such as illegal parking and rubbish exposure, helping the local government to modernize its governance transformation. In addition, the most representative AI-enabled rural revitalization is the Digital Countryside Project launched by Shangtang Technology in Songyang, Zhejiang Province. Through the development of AI courses to promote the development of primary and secondary school studies, AR recovery of traditional ancient villages, rural digital museums, and other rural governance.

With the help of generative AI, art has made significant contributions to rural revitalization and the development of rural areas\cite{zhang2022cgan}. It has increased cultural awareness in regions inhabited by the She ethnic group and attracted numerous designers to re-imagine local ethnic patterns. For example, the She Pattern 2023 International Pattern Creative Design Competition\footnote{Website:https://www.shejijingsai.com/2023/08/994035.html}, launched in August 2023, aims to showcase redesigned ethnic patterns specific to the She culture.


\section{Content Generation Workflow}

\begin{figure}[h]%
\centering
\includegraphics[width=\textwidth]{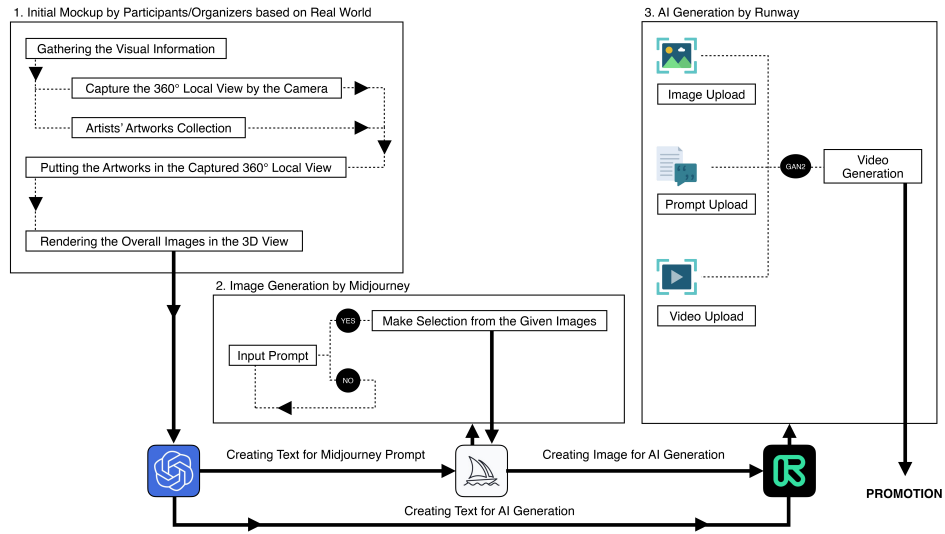}
\caption{AI Content Generation Workflow}\label{fig1}
\end{figure}


The implementation of our system consists of three main stages: initial mockup (3D Rendering), image generation by AI (Midjourney), and AI video generation (Runway). Participants or organizers collect visual information in the initial mockup phase by capturing a 360° local view using a camera and sourcing artwork from artists. These artworks are then digitally placed within the captured local view, rendering the overall images in the 3D environments. This visually rich information serves as input data for a Chat GPT 4.0, which generates text prompts for the subsequent image generation stage.

In the Midjourney stage, which focuses on AI image generation, the prompts generated by the Chat GPT 4.0 are used to select specific images from the 3D environments. If the prompts do not yield suitable images, additional descriptions are added to refine the selection process. The chosen images are then transformed and readied for the final stage, in which AI video generation takes place.

Runway's Generative Adversarial Network 2 (GAN 2) model is employed in the AI video generation stage. The images chosen in the Midjourney phase, the prompts generated by the Chat GPT 4.0 model, and any original video content are used as input for the GAN 2 model. The GAN 2 model is capable of generating video content based on these inputs. Users can further influence the output by providing a style reference or uploading a style image, which guides the AI toward a desired aesthetic in the generated videos. This multi-stage implementation results in a novel blend of real-world views, artistically curated content, and AI-enhanced video output, See Figure~\ref{fig1}.

\section{Methodology}

This study incorporates evidence from a literature review, case studies, semi-structured interviews, and thematic analysis to explore the intersection of complex aesthetics and generative AI and its implications for cultural creativity in rural areas. 

\subsection{Case Study}

We use case studies to explore a specific phenomenon in-depth. This approach yielded detailed insights into the subject matter under investigation. A finite selection of cases was made, and questionnaires and structured interviews were conducted with relevant stakeholders, such as curators, artists, villagers, and visitors. The cases selected for study include the Tongling Country Public Art Festival (2018), Another Wave - the First Daojiao New Art Festival (2016), and the Lianzhou International Foto Festival (2005). These events, each with unique cultural and artistic dimensions, provide a comprehensive and contrasting study landscape. 

Our assessment was based on the festival locations, their international influences, and the fame of the participating artists. Festivals based in more remote or rural areas, such as Tongling, Daojiao, and Lianzhou, have the potential to provide unique cultural experiences and stimulate local development. These locales may not be provincial capitals, but the success of these festivals in garnering attention and fostering cultural engagement has proven that effective planning and implementation, coupled with the power of art, can transform lesser-known places into dynamic cultural destinations. Regarding global impacts, the Lianzhou International Foto Festival has garnered more recognition than the other festivals, largely because it has been an annual event since 2005, whereas the other festivals are relatively new. Another Wave - the First Daojiao New Art Festival was a one-off event.

The participating artists in these festivals hail from diverse backgrounds and enjoy varying degrees of fame. For instance, the Lianzhou International Foto Festival's longstanding history has enabled it to attract a broader network of renowned photographers and artists from around the world, thereby enhancing the festival's global standing and influence, which in turn creates a positive feedback loop of growing recognition and talent attraction. In contrast, the newer festivals, Tongling Country Public Art Festival, and Another Wave - the First Daojiao New Art Festival, may feature a combination of budding and established artists. Their artist lineups may predominantly consist of domestic talent, but this does not detract from their cultural relevance or growth potential. Instead, it may offer a unique platform for local artists to gain exposure and for attendees to be introduced to new artistic perspectives. Despite the one-time occurrence of Another Wave - the First Daojiao New Art Festival, it may have left a lasting impression on the local community and the artists involved. Its sole edition may have acted as a catalyst for future projects or left a significant cultural footprint.

\subsection{Questionnaire}

The setting of the questionnaire was designed to incorporate three question types: apparatus-based questions, five-point scale responses, and open-ended questions. The apparatus-based questions were formulated to categorize interviewees based on their backgrounds. These questions focused on gathering information about the participants' roles, experiences, and qualifications. The survey employed a 5-level scale, where the first level signifies ``strongly disagree" and the fifth indicates ``strongly agree". This scale is designed to aid researchers and decision-makers in discerning the ways and degree to which art activities bolster rural revitalization. Open-ended questions were also included to capture unconsidered details and fill any gaps within the questionnaire. These questions encouraged participants to provide descriptive and nuanced responses, enabling a more comprehensive understanding of their perspectives.

Our questionnaire, with 329 collected responses, served as an exhaustive instrument for collecting data on various crucial aspects such as the festival's contribution to preserving local culture, how the local culture is incorporated into the festival, the significant challenges in organizing the festival, the festival's impact the local community, the audience, and visitors respond to the festival, the role of generative AI in promoting cultural creativity in future festivals. It investigated the respondents' participation in the festival, their viewpoints on the event's main objectives, and their personal standouts. Furthermore, it assessed the festival's influence on the local community and gauged the audience's reaction.

\subsubsection{Data Analyse}

A significant portion of this survey revolved around the interviewer's profound understanding of how generative AI could play a potential role in future iterations of the festival. This perspective was pertinent as most respondents felt that the artworks' interactions did not significantly resonate with the local culture, owing to the perceived isolation of these artworks. About 62\% of respondents viewed the interaction between artists and locals as only marginally significant, rather than as a means of promoting local culture during festivals.

Another prevalent feedback was the economic constraints not ensuring the sustainable growth of art promotion in rural areas. Respondents who identified these effects as severe or significant constituted more than 50\% of all participants. The rationale for this is twofold: the expansion of the pandemic and the demand for artist involvement. We are considering its future potential in applying AI in these rural settings, particularly its utilization by locals. This could reduce the need for artist participation and artwork, thereby streamlining the festival organization.

\subsection{Semi-structured Interview}

We conducted Semi-structured interviews with 24 stakeholders which included artists, curators, and the general audience (Village workers, villagers, village officials.), and 2 curators in-depth interviews. In this study, we also employed thematic analysis to scrutinize data derived from semi-structured interviews. These interviews encompassed a diverse range of 24 participants' interview content. The insights gleaned from these in-depth interviews, when passed through the lens of thematic analysis, offered a profound understanding of the curators' experiences and perspectives, thereby enriching the overall research findings. 

\subsubsection{Participants}

\begin{table}[h]
\caption{Participants' and their information}\label{tab1}%
\begin{tabular}{@{}llll@{}}
\toprule

\textbf{Participants}  & \textbf{Age}  & \textbf{Gender} & \textbf{Year of Experience}\\
\midrule
C1    & 38   & M  & 3  \\
C2    & 34   & F  & 1  \\
C3    & 37   & F  & 1  \\
C4    & 28   & F  & 6  \\
C5    & 31   & M  & 9  \\
A1    & 29   & M  & 2  \\
A2    & 30   & F  & 3  \\
A3    & 44   & F  & 1  \\
A4    & 37   & F  & 2  \\
A5    & 22   & M  & 3  \\
A6    & 34   & M  & 1  \\
A7    & 37   & F  & 1  \\
G1    & 21   & M  & 1  \\
G2    & 24   & M  & 1  \\
G3    & 27   & M  & 2  \\
G4    & 31   & F  & 1  \\
G5    & 44   & F  & 1  \\
G6    & 37   & F  & 0  \\
G7    & 31   & F  & 0  \\
G8    & 24   & M  & 1  \\
G9    & 27   & M  & 0  \\
G10   & 21   & F  & 0  \\
G11   & 24   & F  & 1  \\
G12   & 27   & F  & 2  \\
\botrule
\end{tabular}
\end{table}

Table \ref{tab1} provides information about the participants involved in the study. Here is a summary of the information:
We invite 5 curators, 7 artists, and 12 general audiences. The participants are identified by the codes C1, C2, C3, C4, C5, A1, A2, A3, A4, A5, A6, A7, G1, G2, G3, G4, G5, G6, G7, G8, G9, G10, G11, and G12.
The participants' ages range from 21 to 44 years old ($45\pm1.25$ years).
The participants include males (M) and females (F), representing a mix of genders.
The ``Year of Experience" column indicates the number of years of experience each participant has, ranging from 0 to 9 years.
The table provides a snapshot of the participant's demographic information, including their age, gender, and years of experience.

\subsubsection{In-depth Interviews}

After the Semi-structured Interview with a total of 24 participants, we selected C1 and C3 for in-depth interviews. Their responses in the questionnaire stood out because of their comprehensive insights and unique perspectives on the discussed topics.

The first respondent C1 was chosen because of his detailed insights into incorporating local culture and rural aesthetics into the festival and thoughtful suggestions on how generative AI could promote cultural creativity. In the interview with C1, we focused on his detailed understanding of integrating local culture into the festival and his thoughts on the role of generative AI in promoting cultural creativity. We asked him to elaborate on his challenges and provide further suggestions for improving the use of generative AI in future festivals.

The second respondent C3 was selected due to her comprehensive understanding of the festival's impact on the local community and her forward-thinking ideas on how generative AI could be used to create digital archives and preserve rural culture. For C3, our interview questions were centered around his perception of the festival's impact on the local community and her innovative ideas on how generative AI could be used to create digital archives and preserve rural culture. We sought to understand her experiences with audience and visitor responses and her views on the potential feasibility of using generative AI in upcoming festivals.

From these interviews, we found that both C1 and C3 shared a belief in the potential of generative AI to enhance the representation of local culture and rural aesthetics in the festival. They also highlighted the challenges involved in integrating advanced technology into a traditional setting, such as the need for resources and gaining acceptance from the local community. Their insights provide valuable direction for future research and planning in integrating generative AI into rural art festivals. We delved deeper into these topics, seeking to understand not just the what but also the how and why of their perspectives. This allowed us to better understand the potential for integrating generative AI into future festivals and the possible challenges and opportunities that might arise from this integration.

\subsubsection{Thematic Analysis}

\begin{table}[h]
\caption{Main Themes and relevant codes, and number
 of code occurrences(OC)}\label{tab2}
\begin{tabular*}{1.035\textwidth}{@{\extracolsep\fill}lcccccc}
\toprule%

Theme & Relevant Codes & CO\\

\midrule
From Urban & Enriched creative landscape & 28 \\
to Countryside & Renewed focus on local culture  & 19  \\
 & Sustainable development - Collaboration  & 30  \\
 & Sustainable development - Cultural tourism & 22 \\
 & Sustainable development - Digital platforms  & 15  \\
\toprule & Total & 114 \\
\toprule%
The Form of Art & Reflection of traditions and customs & 45 \\
in Countryside & Celebrating cultural identity & 12  \\
 & Revival of traditional crafts & 32  \\
 & Pottery, paper-cutting, embroidery & 21 \\
 & Influence of landscape and nature & 18  \\
 & Inspiring artistic creations & 22  \\
\toprule & Total & 150 \\
\toprule%
Execution & Enhancing artistic creations  & 39  \\
and Development & Innovative artistic techniques & 26  \\
 & Unconventional approaches & 14 \\
 & Training programs and workshops & 10  \\
 & Knowledge sharing and interdisciplinary exchange & 27  \\
 & Promoting local art and sustainable development & 7  \\
\toprule & Total & 123 \\
\toprule%
Interaction & Community engagement - Input, resources, local knowledge & 25  \\
and Influences & The social impact of art - Cohesion, cultural pride, well-being & 12  \\
 & Broadening perspectives, new ideas & 36 \\
 & Participatory art practices, Co-creation, and community ownership & 23  \\
 & Influence of traditions Inspiring artistic expression & 21  \\
 & Empowerment through art - Self-expression, cultural preservation & 9  \\
\toprule & Total & 126 \\
\toprule%
High-tech & Application of generative AI - Novel artistic outputs & 7  \\
Involvement & Digital technologies for preservation - Digitization, virtual reality & 12  \\
 & Immersive rural settings - Virtual reality experiences & 33 \\
 & Digital platforms - Showcasing, accessibility, cultural exchanges & 39  \\
 & Challenges and opportunities - Infrastructure, affordability, digital inclusion & 17  \\
\toprule & Total & 108 \\
\toprule%
All&   & 621  
\\
\botrule
\end{tabular*}
\end{table}

This study delves into public attitudes towards combining generative AI and local culture in rural art festivals, uncovering diverse viewpoints. Some respondents believe that generative AI can introduce a variety of art and cultural activities to rural areas, while others express skepticism about the feasibility of this integration. There's also a perception among some participants that rural art festivals do not adequately represent local culture and characteristics. Simultaneously, examining rural artistic practices in China has been conducted, focusing on the high frequency of specific keywords in interviews. This analysis suggests that these terms are significant and relevant in the context of rural art. The respondents appear to perceive urban-rural communication as a catalyst for artistic growth, emphasize the importance of preserving local cultural heritage, and seek strategies for sustainable development, as indicated by the frequent use of these keywords. The high frequency of keywords such as``reflection of traditions and customs" and``revival of traditional crafts" suggests that respondents view rural art as a means of preserving and honoring cultural identity. A renewed interest in traditional artistic practices is indicated, potentially spurred by a desire to reconnect with cultural heritage and generate economic opportunities through craftsmanship. The repeated mention of ``enhancing artistic creations" implies a shared aspiration among respondents to improve the quality and impact of rural art. Terms like``innovative artistic techniques,"``training programs and workshops," and ``knowledge sharing and interdisciplinary exchange" indicate a desire to explore new approaches, develop skills, and foster collaboration and cross-disciplinary interactions. Community engagement is frequently mentioned, highlighting its importance in artistic endeavors by involving local communities and leveraging their resources. Respondents also recognize the potential social impact of art, acknowledging its capacity to foster community cohesion, cultural pride, and overall well-being. The limited mention of the``application of generative AI" suggests that the use of artificial intelligence in artistic creation is still relatively nascent in rural contexts. However, using ``digital technologies for preservation" and``digitization, virtual reality" indicates an interest in leveraging digital tools to preserve and present rural art forms.

\section{Key Findings}

The artworks presented by many artists at art festivals are simply moving their pre-existing creations into another``art museum" space rather than creating organic combinations specific to the local context. This approach only benefits the dissemination of art but fails to promote the rural areas' unique characteristics and cultural heritage. To appropriately promote local cultural experiences, creating works organically integrated with the local context is necessary.
Rural areas face a limitation in their revitalization efforts as they rely on artists as external forces who cannot permanently reside in these villages. Hence, rural areas need to cultivate their``artists (villagers)" who better understand the local context and may not necessarily be trained professionals. To this end, generative AI can serve as a tool for these ``artists" as machine learning learns from subjective aesthetics (provided by the previously participated artists) and has the potential to contribute to sustainable development while residing in rural areas. This approach can replace the traditional annual rural art festivals,  and form a sustainable development cycle~\cite{qu2022community}.
There are two ways to use generative AI to protect local cultural heritage, which has yet to be fully explored in rural revitalization. The first is conducting in-depth academic and inheritance-based preservation while respecting the history and cultural heritage relying on the machine learning of generative AI. The second is to attract audiences by providing interactive generated images/texts/videos that offer multiple visual representations, which can promote tourism and rural development while maintaining entertainment value, like digital storytelling~\cite{bogdanovych2010authentic}.

When looking at the relationship between aesthetics, AI, and cultural creativity from a theoretical perspective, it is first necessary to clarify the intrinsic correlation between these. In terms of the concept of complex aesthetics, which corresponds to generative AI. And complex aesthetics reflects generative art, which uses digital media and information interaction as its hub~\cite{arielli2022ai}. Unlike traditional art, which uses physical media and natural language to create, the systematic complexity and typological diversity of complex art reflect different characteristics~\cite{gao2022metanalysis}. The characteristics of complex art can be summarized as three points: 
\begin{itemize}

    \item A greater tendency towards normative language expression, emphasizing the extension of predictable phenomena in both the temporal and spatial dimensions; 
    \item A selective operating mechanism, which sifts out distinguishing and effective data from a large amount of irregular data, which can provide a source of creative consciousness; 
    \item A two-way selectivity of aesthetic value reinforced by the pivot of informational interactions, further narrowing the distance between subjects, which can provide a source of creative awareness; and a two-way selectivity in the interaction of information as a pivot to strengthen aesthetic value, which further reduces the distance between subjects. This further reduces the distance between subjects, which, from the perspective of intersubjectivity, strengthens the construction of the relational network between subjects. To summarize, these three features of complex art highlight the organic, holistic, and processual nature of generative AI. This is compatible with the nature of cultural creativity~\cite{donnarumma2022against}. After all, for promoting cultural creativity, the correspondence between individual creation and the overall cultural environment, the integration of science and humanities, and the exchange of local and diverse cultures are three essential and necessary paths.

\end{itemize}
    Aiming at it, we can think about the possibility of the use of generative AI in art creation to enhance cultural creativity in Rural China from three levels.
    First, cultivating art workers in local cultural contexts. This is due to the fact that artists often have difficulty integrating their artworks with local culture when practicing rural art practices. This problem arises mainly due to the existence of cultural differences. Cultural difference is both a challenge and an opportunity~\cite{stuhr1999celebrating}. On the one hand, China's vast geography makes rural culture have obvious differences between north and south and east and west, and with the development of society and economy, the differences between urban culture and rural culture are more prominent; on the other hand, the existence of cultural differences provides the necessary conditions to stimulate the imagination of artists, and the fusion of aesthetic experiences in multiple contexts can create more universal works of art for artists. In the process of dealing with cultural differences, ``communication" is the most effective method. This exchange is a two-way street. It includes artists' participation in rural art practice, and it also includes local art workers' absorption of external art practice experience, exploration of new ways of artistic expression, and improvement of artistic expression ability. It is worth noting that, for local artists, due to the cultural advantages of the art creation context, they should take the people's artistic expression as the purpose of art creation, focusing on the expression of the aesthetic experience of life.
    Second, scientific cognitivism, as an aesthetic principle, plays an important theoretical guiding role for generative AI to be embedded in the process of aesthetic appreciation, which in turn generates the judgment of aesthetic value. Scientific cognitivism, as an embodiment of a natural aesthetic stance, is the relationship between aesthetics and the environment that Alan Carlson sought to address when he proposed scientific cognitivism in 1979, in which the issue of fact and value is unavoidable. In China, the natural environment of the countryside provides a good natural aesthetic field, in which the appreciator understands and integrates to find aesthetic value in aesthetic experience, change of aesthetic interest, and even psychological healing. Scientific cognitivism can play a role in generative AI embedded in the process of aesthetic appreciation, because this aesthetic principle contains the basic connotation of aesthetic relevance, that is, through different ways of presenting art, the aesthetic object does not have the knowledge, imagery, and ideology associated with the aesthetic experience.
    Thirdly, it injects brand-new cultural stimulation points into a rural culture. In terms of the characteristics of rural culture itself, due to its objective conditions and constraints, it will inevitably appear relatively conservative characteristics. This also makes rural culture characterized by relatively single cultural types, slow updating of cultural expressions, and insufficient motivation for artistic creation. Therefore, in the case of insufficient internal development power, external stimuli are essential for the enhancement of cultural creativity. Generative AI, as a new form of cultural expression, can break through the original boundaries of geography, senses, and other dimensions, and promote the renewal and iteration of the original expression and propaganda of rural culture.

\section{Research Agenda and Limitation}

The research agenda proposed by this study is aimed at pushing the boundaries of generative AI and complex aesthetics to promote cultural creativity within the rural regions of China. This approach seeks to address the economic and cultural challenges faced by these areas due to the aftermath of the COVID-19 pandemic.

\subsection{Research Agenda}

The primary objective of this research is to explore and implement a framework that leverages the intersection of generative AI and complex aesthetics to stimulate cultural creativity in rural China. By feeding machine learning models with subjective aesthetic data, the goal is to enable them to generate output that is culturally creative and reflective of the local aesthetics.
This research also seeks to create a dialogue between technology and culture, hoping to preserve local traditions while embracing the benefits of modernity. In this context, the proposed framework is not meant to replace human creativity but rather to enhance it by infusing it with AI intelligence.
As part of this research agenda, the study invites further exploration and experimentation in the following areas:

\begin{itemize}

\item Development of methodologies for extracting subjective aesthetic data from pre-existing artistic representations.
    
\item Building machine learning models that can effectively learn from and generate output based on this data.

\item Identification of effective ways to weave site-specific artworks into the local context.

\item Exploration of how generative AI can nurture grassroots artists within villages.

\end{itemize}

\subsection{Limitations}

The lack of a standardized method for content quantification makes it challenging to evaluate artworks, creating obstacles in developing rural cultural creativity. Cultivating ``artists (villagers)" is crucial, but preventing subjective bias when training them to use generative AI is challenging. The solution is to let generative AI learn the subjective essence of art from real artists' data and use it to guide villagers to control and implement in real time. Future research could improve generative AI algorithms to better align with cultural experience and local heritage, evaluate its impact on rural communities and cultural heritage, explore new ways to integrate local knowledge, assess social and ethical implications, and compare how generative AI is used in different cultural contexts and rural areas worldwide. Despite the promising potential of this research, there are several limitations to consider:

\textbf{Cultural Creativity:} The subjective nature of aesthetics and cultural creativity may present challenges in defining and extracting subjective aesthetic data that can be processed by machine learning algorithms.

\textbf{Generative AI:} While generative AI has proven effective in various applications, its use in promoting cultural creativity in rural areas is relatively unexplored. As such, there may be unforeseen technical obstacles or limitations to overcome.

\textbf{Data Availability:} The availability and quality of data for both training the machine learning models and evaluating their output may be limited, especially in rural areas.

\textbf{Socio-cultural Acceptance:} There might be resistance to the incorporation of AI and technology in cultural and artistic practices among rural communities. This resistance could stem from fear, misunderstanding, or the desire to maintain traditional practices.

Despite these limitations, the potential benefits of this research in promoting cultural creativity and economic development in rural China should not be underestimated. The proposed framework provides a solid foundation for further research and experimentation in this area.

\section{Conclusion Remark}


In conclusion, this study provides an original perspective on leveraging the intersection of complex aesthetics and generative AI to promote cultural creativity in rural China during the post-pandemic recovery. Through examining applications in cultural heritage and conducting a multi-method analysis, key findings emerged. First, artworks integrated within local contexts are crucial for effectively promoting cultural experiences versus conventional disconnected rural art festivals. Second, nurturing grassroots “artist-villagers” through generative AI aids sustainable development.

Ultimately, this pioneering research proposes a human-centered framework synergizing AI capabilities and nuanced aesthetics to inject new vitality into rural heritage while fostering enlightening dialogues about tradition and modernity. The core contributions are elucidating generative AI's potential to enable cultural creativity, establishing frameworks to extract subjective aesthetic data, and envisioning AI not as a replacement for human creativity but rather as enhancing it. This conclusion provides a foundation for further exploration of the nexus of technology, culture, and creativity.
\\

\bmhead{Acknowledgments}

Acknowledge financial support from the National Social Science Foundation of China (\#20CJY045).


\bibliography{sn-article}

\end{document}